# Magnetoelectric Coupling Based on Protons in Ammonium Sulfate

Lei Meng,[1*] Chen He[1*], Fei Yen,[1†]

[1]State Key Laboratory on Tunable Laser Technology, Ministry of Industry and Information Technology Key Laboratory of Micro-Nano Optoelectronic Information System and the School of Science, Harbin Institute of Technology, Shenzhen, University Town, Shenzhen, Guangdong 518055, P. R. China

**Abstract:** Most ferroelectric crystals have their own set of unique characteristics and ammonium sulfate $(NH_4)_2SO_4$ is no exception. We report on two previously unidentified features in ammonium sulfate: 1) that there are at least two successive transitions instead of one occurring at the Curie temperature $T_C$ = 223 K according to dielectric constant measurements; and 2) pronounced step-like anomalies are found in the magnetic susceptibility $\chi(T)$ exactly at $T_C$. To explain these results, we take into account that there exists a previously unidentified linear coupling between the magnetic and electric dipole moments of the $NH_4^+$ tetrahedra due to their rapid reorientations and distorted geometry, respectively. The magnetic moments are small, 0.0016 $\mu_B$ for every $C_3$ reorientation which involve three protons ($H^+$) undergoing orbital motion. Nevertheless, short-range correlations exist in the paraelectric phase because the magnetic moments are restricted to only point along 14 possible orientations due to the symmetry and periodic nature of the potential wells. At $T_C$, $C_2$ reorientations (involving four protons) are no longer energetically feasible so the reduction in the degrees of freedom to 8 further enhances the effect of the magnetic interactions. This triggers long-range ordering of the orbital moments in an antiferromagnetic configuration along the *ab*-plane, which via Dzyaloshinskii–Moriya interactions, end up canting slightly toward the *c*-axis direction. Since there exists two types of inequivalent $NH_4^+$ groups that reorient at different frequencies with temperature and do not have the same degree of distortion, the emerging polar phase is ferrielectric. This previously unidentified 'magnetoprotonic' effect can be further extended toward understanding the fundamental causes of the spontaneous polarization in many other ferroelectric crystals as well as provide the missing link toward understanding the enhanced functionalities of many hydrogen-based compounds and organic-inorganic hybrid materials.

**Introduction:**

The protons (H⁺) in most hydrogen-based compounds become geometrically ordered at a critical temperature $T_C$. Usually, a structural phase transition accompanies this process and the lattice becomes uniquely distorted. A subset of these systems results in the appearance of ferroelectricity at $T_C$ such as in ammonium sulfate $(NH_4)_2SO_4$,[1] potassium dihydrogen phosphate $KH_2PO_4$ (KDP),[2] even the now trending methylammonium lead bromide $NH_3CH_3PbBr_3$,[3] and the list goes on.[4,5] A common feature in these ferroelectric crystals is that the protons are always in motion and moving charges translate to current. In particular, reorienting ammonium cations $NH_4^+$ carry a magnetic moment because the enclosed areas of the orbits of the positive and negative charges are different so their respective magnetic moments do not cancel each other out (Figs. 1a and 1b). The associated moment is only 0.0016 $\mu_B$,[6] however, due to the shape and symmetry of the periodic potentials, the $NH_4^+$ only exhibit $C_2$ and $C_3$ reorientations so the directions of the moments are restricted to only point along 14 directions (along the diagonals and faces of a cube) which dramatically enhances intermolecular orbital interactions. From such, we concluded that it is the long-range ordering of the proton orbitals in the ammonium halides that trigger their geometric ordering and consequential structural phase transitions. The natural question now is whether long-range ordering of the proton orbitals underlie the observed polarization in hydrogen-based ferroelectrics, if whether they are actually multiferroic all along. We selected ammonium sulfate since the proximities between neighboring $NH_4^+$ of 3.86 Å (Ref. [7]) and the energies associated to the torsional motions of 303–335 cm⁻¹ (Ref. [8]) are extremely similar to that of the already inspected ammonium halides. As we present evidence in this work that the proton orbitals become ordered at $T_C$ in $(NH_4)_2SO_4$, the problem at hand is certainly not limited to magnetic moments generated by reorienting $NH_4^+$. In theory, any reorienting body carries a magnetic moment.[9] For instance, reorientations of $CH_3NH_3^+$, $BH_4^-$ or $CH_3^+$ are also expected to occur in their parent compounds due to the symmetry of the periodic potentials,[10] weak lattice cohesive forces,[11] and zero-point energy. Though yet, researchers typically treat hydrogen atoms as being static because a simultaneous exchange of their positions does not alter the lattice. The torsional frequencies can reach up to $10^{12}$ Hz and depending on the temperature the process is either mainly classical hopping or via quantum tunneling.[12] Hence, by taking into consideration that the orbital motions of protons also carry a magnetic moment and most importantly, that they are confined to only point along a discrete number of directions which enhance their interactions, we may be able to better understand why, for instance, the methylammonium lead halides exhibit such high photovoltaic conversion rates (with some researchers believing that ferroelectricity is the origin);[13] why some borohydrides are extremely efficient in energy storage (with enhanced ionic conductivity in the disordered phases);[14] why certain pharmaceutical drugs are more effective than others;[15] and why only a fraction of $CH_4$ become geometrically ordered in solid methane.[16]

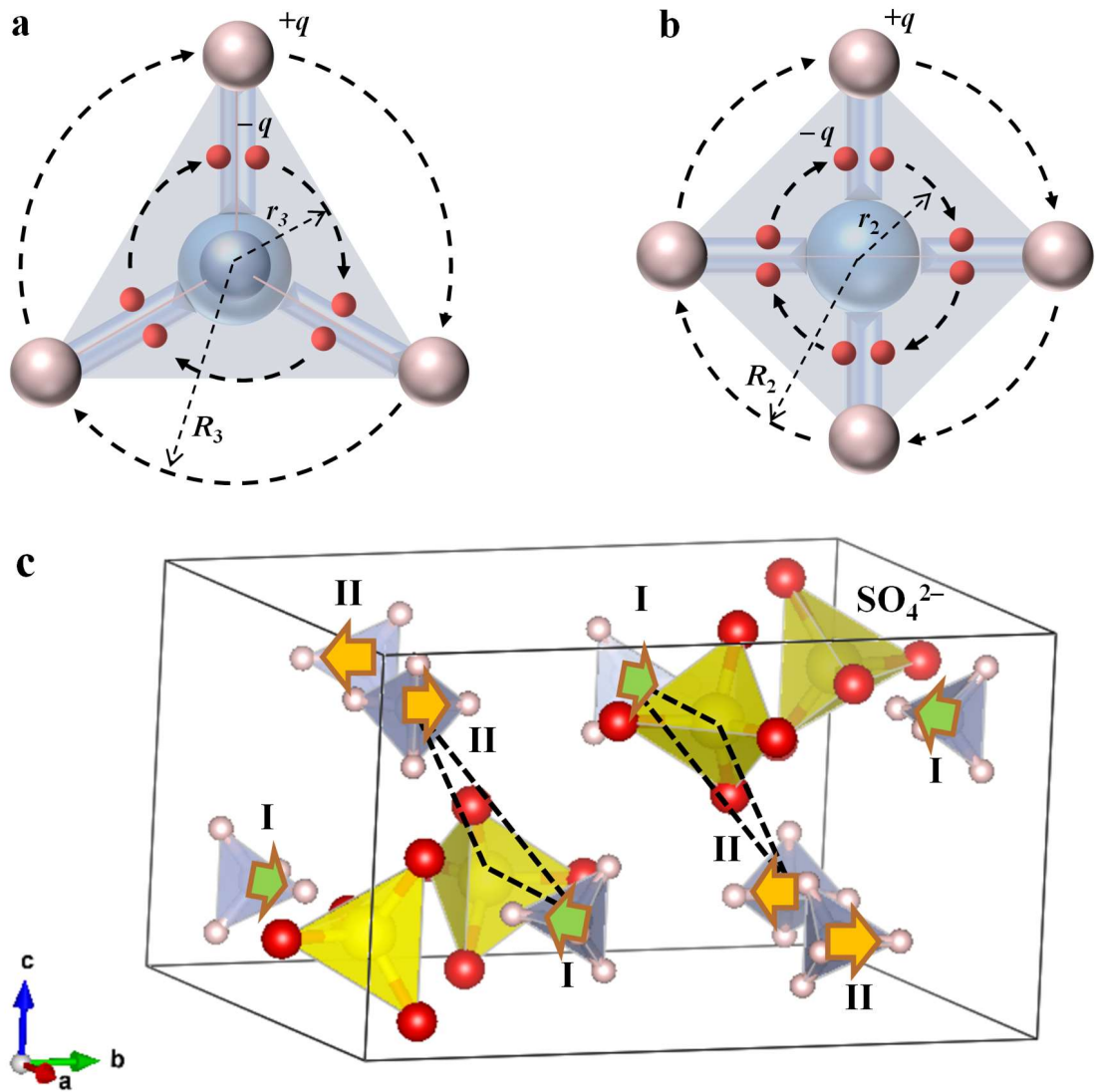

**Figure 1.** The ammonium cation $NH_4^+$ tetrahedron viewed **a**. along one of its vertices and **b**. along one of its edges. When an $NH_4^+$ performs a $C_3$ reorientation, three protons reorient by 120°, this is the equivalent of one proton orbiting one full revolution at a radius of $R_3$. Since protons have a positive charge $+q$, a magnetic moment is generated $\mu_p = (dq/dt)\,\pi R_3^2$. The two electrons linking each proton to the nitrogen atom also reorient with the proton so a magnetic moment in the opposite direction is generated $\mu_e = -2(dq/dt)\,\pi r_3^2$. Since $r_3 < R_3$, and even if $r_3$ is half of $R_3$, $\mu_e$ is half of $\mu_p$ so a residual magnetic moment is always present whenever a tetrahedron reorients by its most elemental step. By the same token, $C_2$ reorientations also generate a residual magnetic moment so the center of each $NH_4^+$ may be treated as a discretized paramagnetic site. **c**. The unit cell of the paraelectric phase of $(NH_4)_2SO_4$ showing the $SO_4^{2-}$ and two inequivalent $NH_4^+$ tetrahedra labelled I and II. Arrows represent the direction of the electric dipole moments. The two types of colored arrows represent the two types of magnetic moments that develop when the system is cooled to $T_C$. The two dashed triangles are expanded in Fig. 6a.

Ammonium sulfate is rather ubiquitous both in the environment and laboratories as it has many commercial uses and is commonly employed as a precipitate in "salting out" and stabilizing proteins.[17] At ambient conditions $(NH_4)_2SO_4$ adopts an orthorhombic motif (Space Group *Pnam*)[18] and when chilled down to $T_C$ = 223 K it undergoes a structural phase transition to the orthorhombic polar Space Group *Pna2₁* with detectable ferroelectricity appearing along the *c*-axis.[1,19] The ferroelectricity is rather peculiar since A) the dielectric constant does not follow the Curie-Weiss law;[20] B) there is an absence of the double hysteresis loops just above $T_C$ which is unusual for first-ordered phase transitions;[21] and C) the polarization exhibits a maximum just below $T_C$ and continuously decreases to even becoming negative below 84.5 K.[22] In addition, a λ-peak is observed in the heat capacity at $T_C$,[20,23] which is usually a manifestation of a breaking of the time-reversal symmetry. Several dozens of papers abound regarding the origin of ferroelectricity in $(NH_4)_2SO_4$ alone. Neutron diffraction measurements[7] concluded that the $NH_4^+$ tetrahedra tilt away from the *ab*-plane upon entering the ferroelectric phase. NMR[24] and ESR[25] measurements observed the $NH_4^+$ tetrahedra to become even more distorted in the ferroelectric phase. On the other hand, some reports suggest the ferroelectricity to be of the order-disorder type where the $NH_4^+$ oscillate between two sites that are 30° about the *ab*-plane in the paraelectric phase which then "freezes-in" in the ferroelectric phase to reside in one of the valleys of a double-well potential.[21,24,26] More recently, researchers have identified that the $SO_4^{2-}$ anions also play a role in the ferroelectricity.[27,28] Interestingly, there exists two inequivalent $NH_4^+$ sites in $(NH_4)_2SO_4$ according to neutron diffraction,[7] deuteron magnetic resonance[24] and spin-lattice relaxation time measurements.[29] These two sites are not only slightly geometrically different (defined by their proximity to the $SO_4^{2-}$), but they reorient at different frequencies which, according to our model, lead to two magnetic moments with distinct magnitudes. To further complicate matters, the $SO_4^{2-}$ also reorient, albeit at much lower rates. The two types of $NH_4^+$ and $SO_4^{2-}$ tetrahedra are also distorted in both the para- and ferro-electric phases rendering them to possess intrinsic dipole moments. Henceforth, if indeed long-range ordering of the proton orbitals occurs at $T_C$, then there should be a simultaneous breaking of the time-reversal and spatial-inversion symmetries (as shown in Fig. 1c) which would explain the appearance of spontaneous polarization. To verify if whether the fundamental cause of the geometric transformation of $NH_4^+$ is magnetic in nature, we measured the *dc* magnetic susceptibilities along the three principle axes of the crystal near the transition temperature and discuss a possible scenario on how and why $T_C$ occurs.

**Methods:**

Ammonium sulfate of 99.997% purity was acquired from Aladdin, Inc. Shanghai, China. Seed crystals of $(NH_4)_2SO_4$ were grown from the slow evaporation solution growth method using deionized water. Samples varied from 12 to 40 mg in weight. The *dc* magnetic susceptibility was measured using the vibrating sample magnetometer (VSM) of a Physical Properties Measurement System (PPMS), Dynacool series unit manufactured by Quantum Design, U.S.A. GE varnish was used to mount the crystals onto quartz palettes or copper rods for the measurements. Cooling and warming speeds near the transition temperature were typically set to 0.5-1 K/min. No electric field bias was applied to any of the samples. The orientations of the samples were identified from their growth habits and double checked with dielectric constant measurements. The real and imaginary parts of the dielectric constant, respectively, were derived from the measured capacitance and loss tangent by an Agilent E4980A LCR meter at 1 kHz sweeping temperature 0.1 K/min. All measured samples were transparent but became bleached after each run past below the transition temperature due to microcracking. The lattice structures were drawn and analyzed with the freeware VESTA (Visualization for Electronic and STructural Analysis) Ver. 3.4.7.[30]

**Results and Discussion:**

Before presenting the magnetic susceptibility results, we show the real $\varepsilon'(T)$ and imaginary $\varepsilon''(T)$ parts of the dielectric constant of $(NH_4)_2SO_4$ at 1 kHz near $T_C$ (Fig. 2a and 2b) because we identify the presence of two peaks very close to each other in temperature instead of only one. During cooling the peaks are situated at 222.30 K and 222.42 K (inset of Fig. 2a) while during warming at 223.58 K and 223.83. Hoshino et al. also reported on the presence of two successive peaks in the heat capacity but did not provide an explanation.[20] Moreover, two additional abrupt changes in the slope of $\varepsilon'(T)$ found above and below the two peaks indicate that the phase transition is at least a two-step process involving the negotiation of competing interactions. The discontinuities also manifest at the same critical temperatures in $\varepsilon''(T)$ (inset of Fig. 2b). When the sweeping rates are larger than 1 K/min, only one peak is observed and the results are in good agreement with those reported by other researchers.[20,21,31-35]

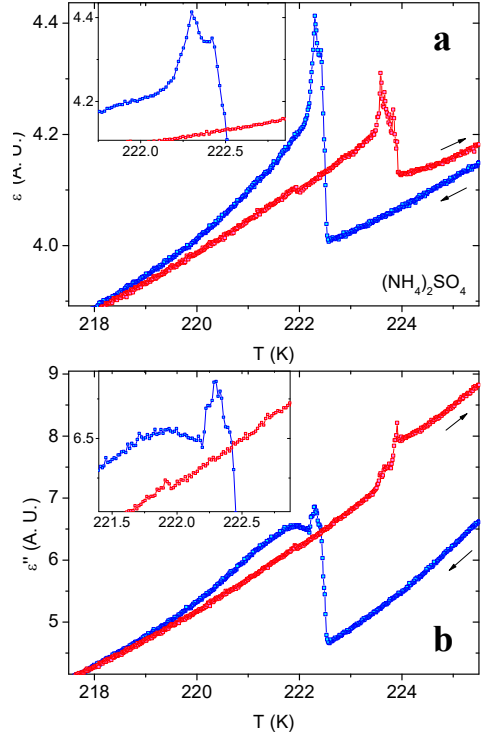

**Figure 2: a.** Real $\varepsilon'(T)$ and **b.** imaginary $\varepsilon''(T)$ parts of the dielectric constant at 1 kHz during cooling and warming near $T_C$ at sweeping rates of 0.1 K/min. Insets are enlarged regions near the transition during cooling.

Figure 3a shows the magnetic susceptibility $\chi_b(T)$ with respect to temperature measured along the *b*-axis orientation of $(NH_4)_2SO_4$ under an external magnetic field of $H$=10 kOe. A sharp step-down anomaly occurred at $T'_{C\_b}$=221.9 K during cooling and the reverse was observed during warming with a step-up discontinuity at $T''_{C\_b}$=224.2 K. When the susceptibility and field was applied along the *a*-axis, similar features were observed in $\chi_a(T)$ with $T'_{C\_a}$=222.2 K and $T''_{C\_a}$=224.2 K (Fig. 3b). For the case of $\chi_c(T)$, the features reversed into a step-up and step-down anomaly during cooling and warming, respectively, at $T'_{C\_c}$=221.5 K and $T''_{C\_c}$=224.8 K (Fig. 4a). The change in the susceptibility along this direction was also the largest comprising almost 3%. The anomalies remained fairly unchanged even when $H$=90 kOe for all orientations; Figure 4b shows the case for when $H$ was applied along the *c*-axis. All discontinuities are sharp and possess a hysteresis of 2-3 K indicative of a first-order phase transition. The observed critical temperatures are in excellent agreement with dielectric[20,21,31-35] and heat capacity measurements[20,23,36] where the transition temperature has been reported to occur in between 218-225 K. We also employed a SQuID magnetometer to measure $\chi(T)$ of crystallites (around 2 mg in weight) straight from the reagent bottle and obtained the same results (shown in the Supporting Information section). The fact that a presence of any non-diamagnetic impurities would easily mask the observed anomalies (especially at 90 kOe) along with the observed magnetic anisotropy leads us to conclude that the observed features are intrinsic of the system. We emphasize that discontinuities do not appear in $\chi(T)$ at solid-solid structural phase transitions if there is no magnetic ordering involved. This holds true even at the melting temperature. In the present case, the change in $\chi(T)$ is small because the magnetic moments of reorienting $NH_4^+$ are only a small fraction of that of electrons. An applied field can only induce so much change to these moments which is what is ultimately measured in the magnetization. The results for when $H = 90$ kOe are also explainable since we are not expecting the proton moments to be easily influenced by $H$ because of apparent strong proton orbital-orbital interactions. From such, we regard these discontinuities as strong evidence that the ordering of the proton orbitals at $T_C$ is magnetic in nature. Furthermore, in our previous study of ammonium iodide[6] and hexamethylbenzene,[37-39] only a step-up anomaly was observed upon entering the proton ordered phases. This could be interpreted as a disruption of the electrons' Larmor precession responses to $H$ due to the ordering of the proton orbitals. However, a step-down anomaly in the present case suggests the proton orbitals first align along the *ab*-plane which then cant toward the *c*-axis. This further substantiates that reorienting $NH_4^+$ possess finite orbital magnetic moments that interact with each other in $(NH_4)_2SO_4$.

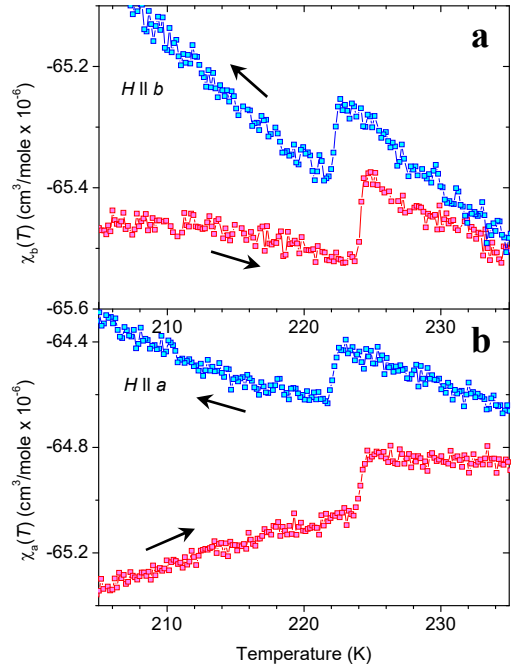

**Figure 3.** Molar magnetic susceptibility $\chi_b(T)$ of $(NH_4)_2SO_4$ along the **a.** *b*-axis and **b.** *a*-axis directions under external magnetic field of $H$=10 kOe.

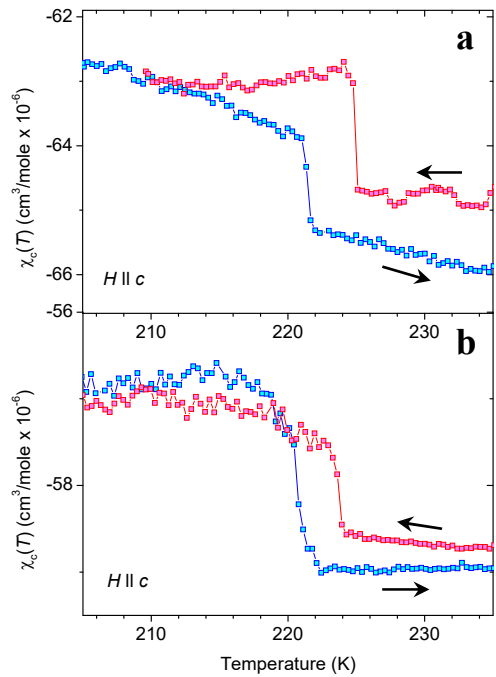

**Figure 4.** Molar magnetic susceptibility $\chi_c(T)$ of $(NH_4)_2SO_4$ along the *c*-axis for when **a.** $H$=10 kOe and **b.** 90 kOe.

Magnetic field dependent measurements of the magnetization *M(H)* were also carried out for all of the three axes above and below $T_C$. The *M(H)* curves along all orientations were linear and negative both above and below $T_C$ so only the case for when $H$ was applied parallel to the $c$-axis is shown in Fig. 5. No hysteretic regions were observed in the $M(H)$ curves confirming that the cooling and warming curves not retracing each other is not due to ferromagnetic ordering. Reports on other physical parameters of $(NH_4)_2SO_4$ usually show only the warming curve. In the few cases where both the cooling and warming curves were displayed,[33-36] they did not retrace each other. This is explainable since the sample cracks into many pieces when cooled past $T_C$ and the piezoelectric effects stemming from the different domains render the system to exhibit plastic deformations as reported in Ref. [40]. The fact that we observe mismatches in the cooling and warming curves of $\chi(T)$ indicate that the spontaneous strain[20] involved may also be magnetic in origin.

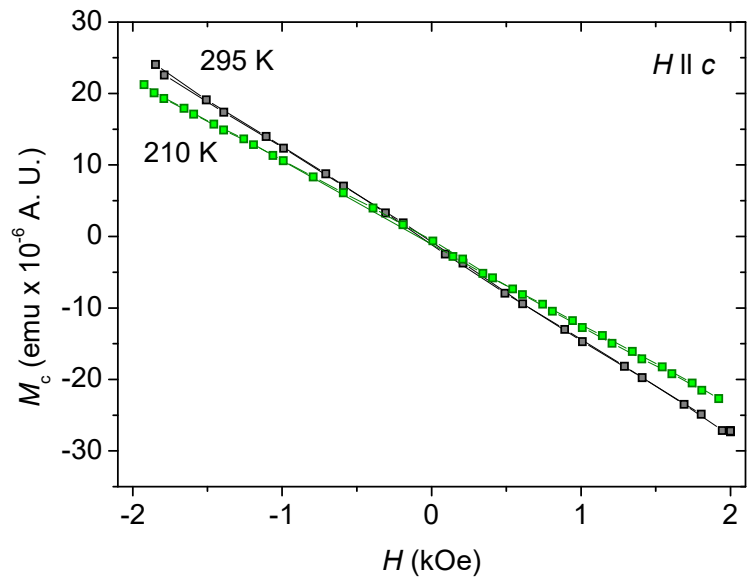

**Figure 5.** Magnetization $M$ versus applied magnetic field $H$ at temperatures above and below $T_C$.

If all of the $NH_4^+$ in the lattice were perfect tetrahedra and rotated along the same direction and at the same frequency, then they would resonate at each others' natural frequencies and the crystal would shatter. To avoid this scenario, adjacent $NH_4^+$ have to orient along different directions in order to evenly distribute the resonant forces. There are 6 possible ways for $NH_4^+$ to exhibit a $C_2$ reorientation: along the clockwise and counter-clockwise directions about each of the three principle axes (Fig. 1b). For

$C_3$ reorientations, there are 8 possible ways: clockwise and counter-clockwise directions about the normal of each of the four sides (Fig. 1a). If the extent of the magnetic interactions were limited to nearest neighboring sites, then the lattice could afford to allow all of its $NH_4^+$ to reorient randomly with the only proviso that no two adjacent $NH_4^+$ reorient along the same direction. Next nearest neighbors can reorient along the same direction because they would be too far away to interact with each other. However, if the interactions extended further than next nearest neighbors, then the 14 possible types of reorientations would no longer be sufficient. This causes some of the tetrahedra to have to distort in accords to their local field. In other words, no two rotating bodies can possess the same energy within the same sphere of influence so any energy degeneracy must be "lifted", a sort of Jahn-Teller effect but applied to protons. In the paraelectric phase, the $NH_4^+$ and $SO_4^{2-}$ tetrahedra are already distorted so short-range correlations appear to already be present. At $T_C$, not enough thermal energy is available to propel 4 protons to exact $C_2$ reorientations but enough energy remains available to allow 3 protons to exhibit $C_3$ reorientations. This presumption is taken from the notion that the reorientation of $NH_4^+$ in ammonium-based compounds are mainly $C_3$ at low temperatures.[41] A more visually apparent example can be found in the reorientations of $CH_3NH_3^+$ in methylammonium lead halides where they not only rotate about the C–N axis ($C_3$) but also exhibit 'tumbling' ($C_4$) at room temperature. The highly energetic $C_4$ motions cannot be sustained upon cooling so only the $C_3$ reorientations remain in the low temperature orthorhombic phase.[42] Coincidentally, existence of ferroelectricity in $CH_3NH_3PbBr_3$ was recently reported in the tetragonal and orthorhombic phases.[43] Reverting to the problem at hand, upon cooling past $T_C$, the reduction of 14 possible ways to reorient to only 8 appear to cause long-range ordering of the reorientations and tilting of the $NH_4^+$ tetrahedra further elaborated below.

The direction of the magnetic moments should align with the long-axis of distorted tetrahedra since it is most energetically favorable to rotate a body about its shortest dimension. This means that the electric and magnetic dipole moments are intimately coupled: aligning parallel or antiparallel with respect to one another depending whether the rotations are clockwise or counterclockwise. As shown in Figure 1c, the electric dipole moments of the $NH_4^+$(I) and $NH_4^+$(II) groups in the paraelectric phase at room temperature reside mainly along the *ab*-plane and cancel each other out. Upon reaching $T_C$, it just so happens that the most stable magnetic configuration to arise is also that of shown in Fig. 1c where the arrows can now represent both the electric and magnetic dipole moments. This configuration was concocted by first assigning the magnetic moments between the two closest $NH_4^+$ (3.859 Å) as being antiparallel to each other followed by the next closest pair and so on. Hence, when mainly $C_3$ reorientations dominate, guided by the dipole moments, the magnetic moments eventually collapse to only exist in two states: the first one shown in Fig. 1c and the other with all of the arrows pointing in the opposite direction which amplifies yet again the effect of the magnetic interactions. To reach to this final configuration, the moments have to find their way staring from an 8-fold state which explains the series of transitions observed in $\varepsilon'(T)$ and $\varepsilon''(T)$ (Fig. 2). The magnetic ordering is inhomogeneous and spatially invariant since the directions and magnitudes of the moments change upon inversion of the three coordinates so macroscopic polarization is to be expected.[44,45] Cheong & Mostovoy provide an excellent summary of many types of systems that can develop macroscopic polarization simply from spatial invariance of the ordered magnetic configuration.[45] In the present case, the dashed triangles in Fig. 1c (expanded in Fig. 6a) show one of many plausible $NH_4^+$(I) – $SO_4^{2-}$ – $NH_4^+$(II) antisymmetric-exchange interactions which causes the magnetic moments of the two ammonium cations to have the tendency to further cant away from the already augmented antiferromagnetic configuration. This involves the two magnetic moments $S_1$ and $S_2$ to tilt out of the *ab*-plane to give rise to a *c*-axis component. This interpretation is in agreement with the step-down anomalies in $\chi(T)|_a$ and $\chi(T)|_b$ as it reflects the spins slightly rotating out from the *ab*-plane and onto the *c*-axis where a step-up anomaly is observed in $\chi(T)|_c$. Note that if the magnetic moments stemmed from electrons, the step-down discontinuities would be step-up and vice versa.[46] Since the magnetic moments of $NH_4^+$(I) and $NH_4^+$(II) are different in magnitude, the degrees of canting are different for the two groups so a net polarization along the *c*-axis arises (Fig. 6b) because, as mentioned above, the electric moments are linearly coupled to the magnetic moments. Moreover, the degree of distortion of the two $NH_4^+$ groups are also different.[47] According to O'Reilly & Tsang the correlation times τ for rotation of $NH_4^+$(I) is slightly smaller than that of $NH_4^+$(II) above $T_C$.[24] However, just below $T_C$, τ of $NH_4^+$(I) and $NH_4^+$(II) increase by 40% and 15%, respectively, rendering the former to reorient slower than the latter. With decreasing temperature, the difference in τ between the two increases. This explains why the polarization begins to decreases continuously just below $T_C$ and even becomes negative at 84.5 K: initially the average tilt angle of the $NH_4^+$(I) tetrahedra is greater than that of the $NH_4^+$(II) because the magnitude of the magnetic moments are larger so a net spontaneous polarization points along the "+" *c*-axis direction, with decreasing temperature, because the rotations of $NH_4^+$(I) decrease much faster than that of $NH_4^+$(II), the tilt angle away from the *ab*-plane of the former decreases faster to eventually reverse the sign of the polarization.

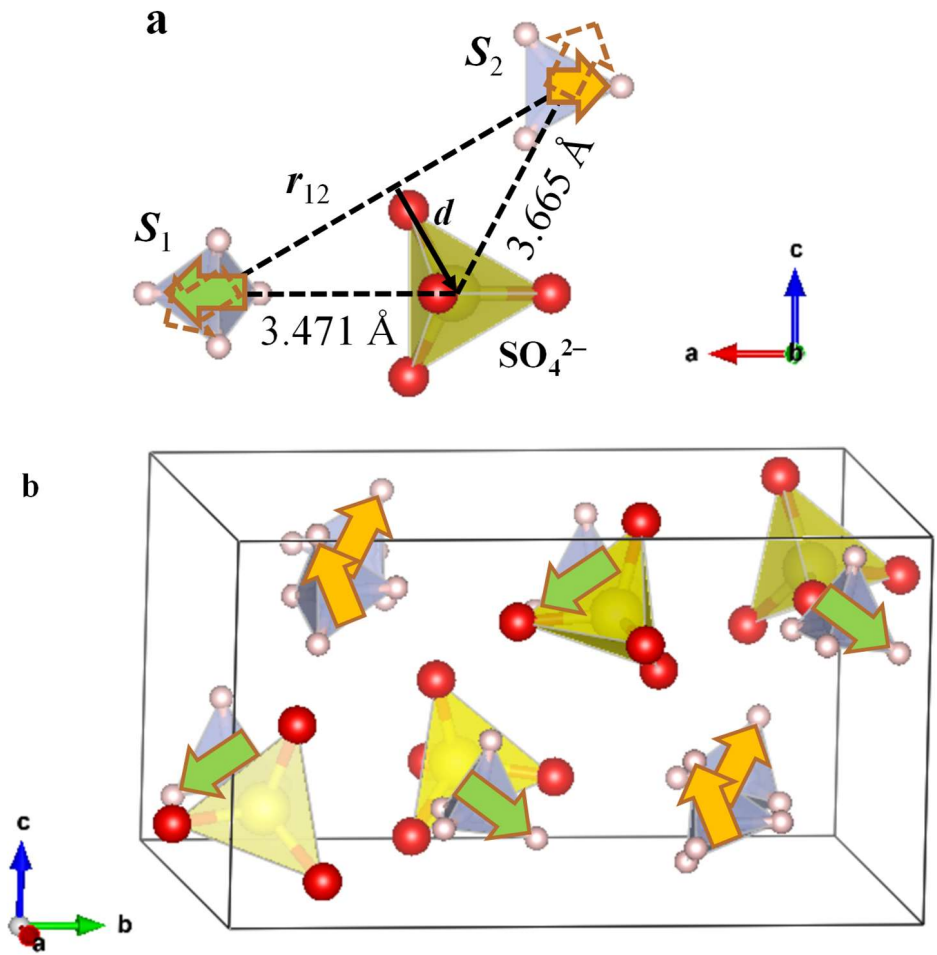

**Figure 6. a.** Antisymmetric exchange type of interaction in $(NH_4)_2SO_4$ between the two magnetic moments $S_1$ and $S_2$ via an $SO_4^{2-}$ ion. The end effect is the canting of $S_1$ and $S_2$ about the Dzyaloshinskii vector $D_{12} = d \times r_{12}$ to a new position (dashed arrows). Since $S_1$ and $S_2$ are different in magnitude and the electric and magnetic dipole moments are linearly coupled, a ferrielectric configuration emerges. **b.** Addition of all of the exchange interactions yield the resultant ferrielectric phase comprised of two groups of dipole moments that do not cancel each other out rendering the system to possess a net polarization along the *c*-axis.

**Conclusions:**

To conclude, $NH_4^+$ exhibiting $C_2$ and $C_3$ reorientations may be regarded as weak paramagnetic sites in the paraelectric phase ($T > T_C$), however, since the directions of the moments are confined to only point along 14 possible directions, short-range correlations exist. At $T_C$, the decrease in the degrees of freedom further enhances the effect of the magnetic interactions between $NH_4^+$ allowing for long-range order. The transition to the magnetic configuration with the lowest energy is not immediate because there exist other configurations with slightly higher energies which explains the cascade of anomalies observed in the dielectric constant measurements before arriving to the most stable spin configuration shown in Fig. 1c. A similar effect was observed in hexamethylbenzene where four pronounced anomalies were found to occur within a range of 3 K at its low temperature structural phase transition.[48] Since two inequivalent magnetic sites exist in $(NH_4)_2SO_4$, the canting of the moments along the *c*-axis via antisymmetric exchange interactions give rise to the famed ferrielectric phase. We note that we also observed magnetic anomalies occurring exactly at the Curie temperatures of potassium dihydrogen phosphate $KH_2PO_4$ and ammonium dihydrogen phosphate $NH_4H_2PO_4$ in our preliminary results; the former is ferroelectric below 122 K and the latter antiferroelectric below 148 K.[2,49] At their respective transition temperatures, the protons of the $H_2PO_4^-$ anions become ordered[50] so their ferroelectric phases also appear to be magnetically-driven. It appears that the spontaneous polarization arising from such hydrogen-bonded ferroelectrics is due to strongly correlated motion of protons all while the electrons play a lesser role. Apart from the observed 'magneto-protonic' coupling, other proton-based parallels such as magneto-optic, magneto-elastic and thermoelectric effects should also exist which may provide the missing link to understanding the photovoltaic properties of organic-inorganic hybrid perovskites and ionic conductivities of certain borohydrides.

**Acknowledgements:**

**Associated Content:**

Supporting Information: Figures S1 and S2.

**Author Information:**

*L. Meng and C. He contributed equally.
Corresponding Author:
†F. Yen, E-mail: fyen@hit.edu.cn. Tel: +86-1334-929-0010.
ORCID: 0000-0003-2295-3040
Notes:
The authors declare no competing financial interests.

**Table of Contents Graphic:**

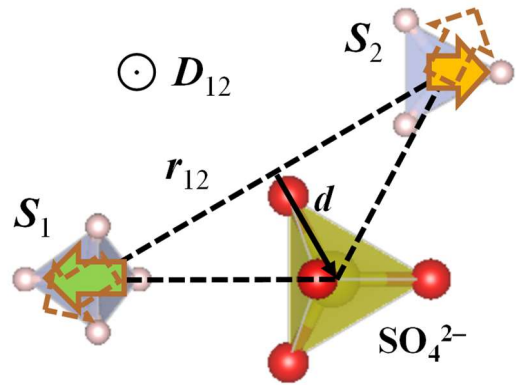

# Magnetoelectric Coupling Based on Protons in Ammonium Sulfate

Lei Meng,[1*] Chen He[1*], Fei Yen,[1†]

[1]State Key Laboratory on Tunable Laser Technology, Ministry of Industry and Information Technology Key Laboratory of Micro-Nano Optoelectronic Information System and the School of Science, Harbin Institute of Technology, Shenzhen, University Town, Shenzhen, Guangdong 518055, P. R. China

**Supporting Information:**

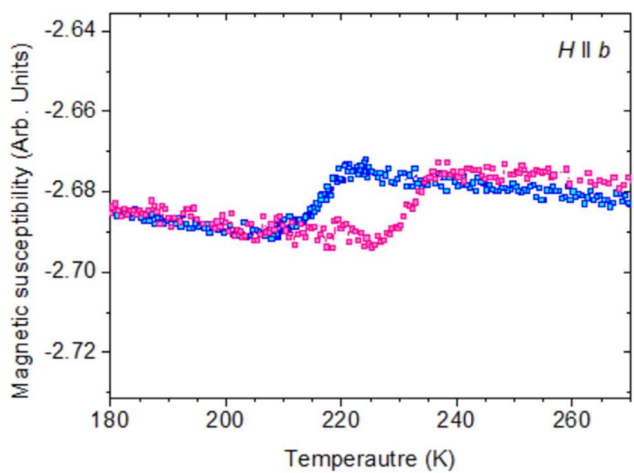

**Figure S1.** Magnetic susceptibility of an ammonium sulfate $(NH_4)_2SO_4$ crystallite (2.1 mg) with respect to temperature under $H$ = 10 kOe aligned along the $b$-axis with an MPMS (Magnetic Properties Measurement System) SQuID magnetometer manufactured by Quantum Design, U.S.A.

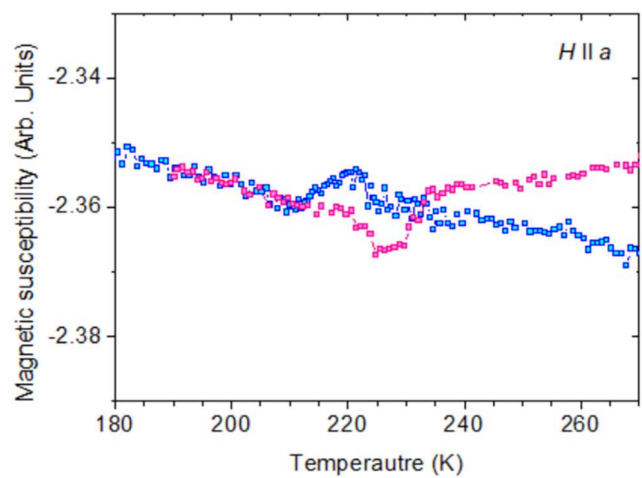

**Figure S2.** Magnetic susceptibility of the same $(NH_4)_2SO_4$ sample also under $H = 10$ kOe but oriented along the *a*-axis.

The obtained results with the MPMS magnetometer were nearly the same. One exception was that the hysteresis regions were larger; this is because the sweeping rates were set to 1.5 K/min instead of 0.5 K/min when using the PPMS as shown in Figures 3a, 3b, 4a and 4b.